# Tau Tagging at Atlas and CMS

Giuseppe Bagliesi

*Istituto Nazionale di Fisica Nucleare – Sezione di Pisa*
*E-mail: Giuseppe.Bagliesi@pi.infn.it*

**Abstract.** The tau identification and reconstruction algorithms developed for the LHC experiments Atlas and CMS are presented. Reconstruction methods suitable for use at High Level Trigger and off-line are described in detail

**Keywords:** tau, HLT, trigger, tagging, Atlas, CMS

**PACS:** 14.60.Fg, 14.60.-z

## INTRODUCTION

A very good identification of the leptons is a basic ingredient of many analyses at the Large Hadron Collider. In particular τ-leptons, which are the most difficult leptons to identify, are expected to be produced by the decay of several interesting physic channels, like Higgs (h/H/A→ττ and H$^{\pm}$→ τν), SUSY and other exotic particles decays. It has been shown (see Figure 1, Ref. 1,2,3) that in a large range of the parameter space, τ identification is very effective in discarding the background (which is mainly due to QCD jets), keeping a good efficiency for signal. The most interesting and peculiar decays are fully hadronic τ decays (τ-jets) since leptonic τ decays are usually identified through the muon or electron produced. In this paper we will concentrate on the methods developed by Atlas and CMS to identify τ-jets.

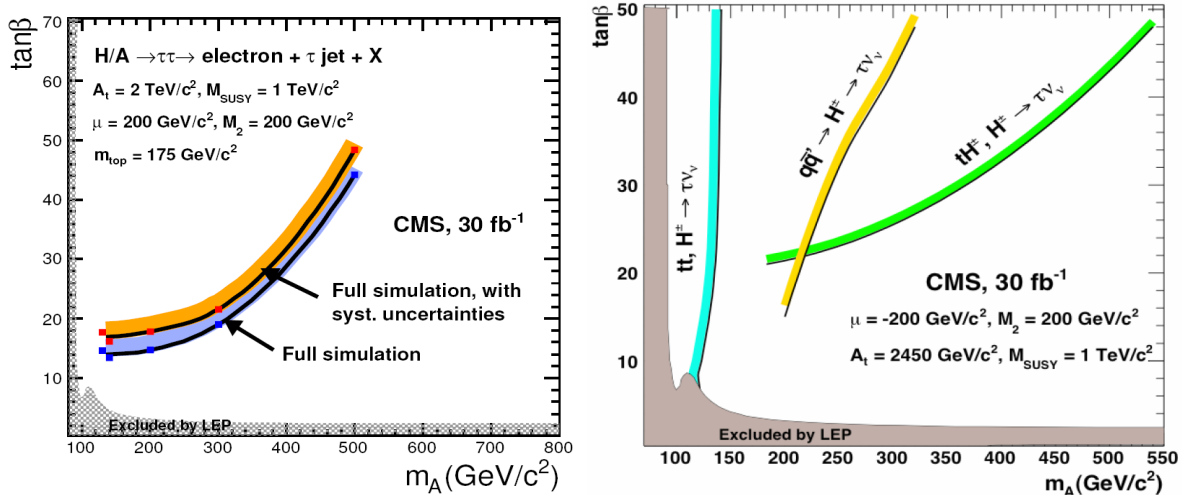

**Figure 1** CMS 5σ discovery region in the $M_A$-tan(β) plane for an integrated luminosity of 30 fb$^{-1}$ in the $m_h^{max}$ scenario.
Left plot: H/A→ττ→ e + τ-jet + X with and without the effect of background systematic taken into account.
Right plot: H$^{\pm}$ → τν →τ-jet + ν + X, for various production mechanisms. The regions already excluded by LEP are shown.

# Properties of the τ leptons at LHC

The τ lepton decays hadronically with a probability of ~65%, producing a τ-jet containing a small number of charged and neutral hadrons. When the momentum of the τ is large compared to the mass a very collimated jet is produced (see Figure 2). For example for a transverse momentum $P_T>50$ GeV/c, 90% of the energy is contained in a cone of radius $R = \sqrt{(\Delta\eta)^2 + (\Delta\varphi)^2} = 0.2$. Hadronic τ decays have low charged track multiplicity (one or three prongs) and a relevant fraction of electromagnetic energy deposition in the calorimeters due to photons coming from the decay of neutral pions. In Table 1 the main τ-decays branching ratios are shown. Quite often taus are produced in pair (like the decay h/H/A→ττ) : in this case 42% of the final states will contain two τ-jets.

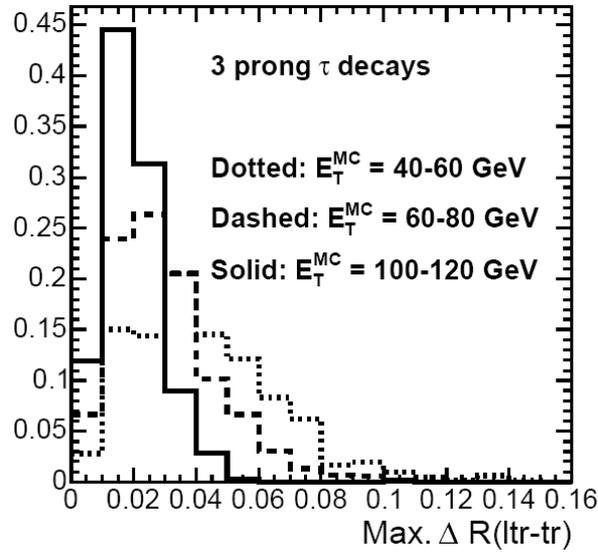

**Figure 2** Maximal distance ΔR in η−φ space between the leading $p_T$ charged particle and the other two charged particles in the three-prong τ decay for three intervals of the true τ-jet transverse energy $E_T^{MC}$

| | |
|---|---|
| $\tau \to e\nu_e\,\nu_\tau,$ | 17.8 % |
| $\tau \to \mu\nu_\mu\,\nu_\tau$ | 17.4 % |
| $\tau \to \pi^\pm\nu_\tau$ | 11.1 % |
| $\tau \to \pi^0\pi^\pm\nu_\tau$ | 25.4 % |
| $\tau \to \pi^0\pi^0\pi^\pm\nu_\tau$ | 9.19 % |
| $\tau \to \pi^0\pi^0\pi^0\pi^\pm\nu_\tau$ | 1.08 % |
| $\tau \to \pi^\pm\pi^\pm\pi^\pm\nu_\tau$ | 8.98 % |
| $\tau \to \pi^0\pi^\pm\pi^\pm\pi^\pm\nu_\tau$ | 4.30 % |
| $\tau \to \pi^0\pi^0\pi^\pm\pi^\pm\pi^\pm\nu_\tau$ | 0.50 % |
| $\tau \to \pi^0\pi^0\pi^0\pi^\pm\pi^\pm\pi^\pm\nu_\tau$ | 0.11 % |
| $\tau \to K^\pm X\nu_\tau$ | 3.74 % |
| $\tau \to (\pi^0)\pi^\pm\pi^\pm\pi^\pm\pi^\pm\pi^\pm\nu_\tau$ | 0.10 % |
| others | 0.03 % |

**Table 1** Most relevant τ decay branching ratios

# Identification of hadronic τ decays: general ideas

Atlas and CMS have developed dedicated algorithms for the identification of τ-jets. In this section the basic, common ideas will be explained.

*Calorimetric isolation and shape variables*

Hadronic τ decays produce localized energy deposit in the electromagnetic and hadronic calorimeters. To exploit this characteristic several isolation parameters which give a measurement of the energy in a ring around the core of the jet have been defined : real taus are expected to release only a small fraction of energy in this ring.

Atlas defines a variable $\Delta E_T^{12} = \sum_{j=1}^{n'} E_{Tj} / \sum_{i=1}^{n} E_{Ti}$ where the sum in the numerator runs over all the calorimeter cells in the cluster with 0.1<ΔR<0.2 respect to the jet direction and the sum in the denominator runs over all the cells with ΔR<0.4. A similar variable is defined by CMS.: $P_{ISOL} = \sum_{\Delta R<0.40} E_T - \sum_{\Delta R<0.13} E_T$ . In Figure 3 the performance of the $P_{ISOL}$ cut are shown for tau and QCD events.

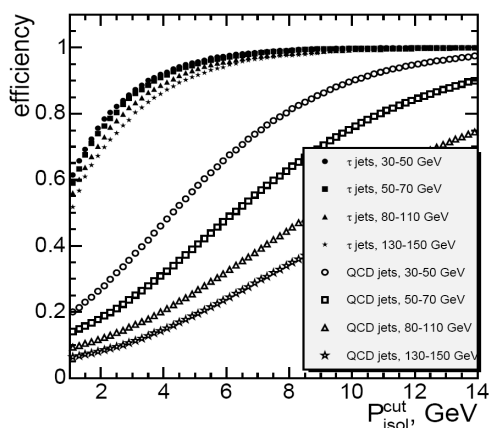

**Figure 3:** The efficiency of the electromagnetic isolation for τ jets and QCD jets in several intervals of the true transverse energy when the value of the cut $P_{ISOL}$ is varied.

Another interesting shape variable is the EM-Radius (see Ref. 4) which exploits the smaller transverse profile of the τ- jets.

*Charged track isolation*

The few and collimated charged tracks contained in a τ-jet are the basic ingredient of a powerful selection algorithm based on the isolation. The principle is shown in Figure 4. The direction of the τ-jet is defined by the axis of the calorimeter jet. The tracks above a threshold of $P_T^{min}$ and in a matching cone of radius $R_m$ around the calorimeter jet direction are considered in the search for signal tracks. The leading track ($tr_1$) is defined as the track with the highest $P_T$. Any other track in the narrow signal cone $R_S$ around $tr_1$ and with z-impact parameter $z_{tr}$ close to the z-impact parameter of the leading track is assumed to come from the τ decay. Tracks with $\Delta z_{tr}$ (impact parameter distance from the leading track) smaller than a given cut-off and transverse momentum above a threshold of $p_T^i$ are then reconstructed inside a larger cone of the size $R_i$. If no tracks are found in $R_i$ cone, except for the ones which are already in the $R_s$ cone, the isolation criteria is fulfilled.

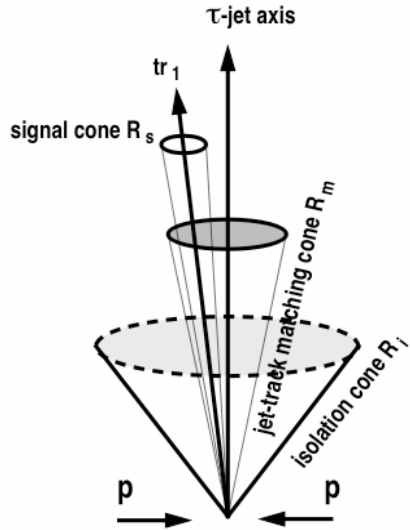

**Figure 4** Sketch of the basic principle of τ-jet identification using the tracker isolation

*Other tau characteristics suitable for tagging*

In addition to the calorimetric and tracker isolation τ lepton has other peculiarities that can be used successfully for tagging. In this section the most important ones are described:

***Number of tracks:*** Hadronic taus decay into 1 (49.5%), 3 (15.2%) and very seldom 5 (0.1%) charged particles plus eventually $\pi^0$. A tagging criterion is to define a track association algorithm which identifies the track belonging to the τ-jet (like a cut on impact parameter ($\Delta z_{imp}$) or in $\Delta R$), and then require that precisely one or three tracks are associated. This cut can be either enforced by adding the condition that the *total charge* has to be ±1, or loosened by asking for a maximum of three tracks (in order to take into account possible track reconstruction inefficiencies).

***Lifetime:*** τ lifetime ($c\tau=87\mu m$) and relatively low mass ($m_\tau=1.78$ Gev/c$^2$) produce a sizeable decay length at the energies of interest for LHC analyses. The decay path allows for the reconstruction of the decay vertex of three and five-prong decays. However since the tracks are very collimated the reconstruction of the decay vertex poses a challenge: a big number of hits are shared in the vertex detectors, which can lead to a reconstruction of fake vertices (see Ref. 6, for example). In the plane transverse to the τ-jet axis the resolution of the reconstructed decay vertex is ~20-30μm. In the direction parallel to the τ-jet axis the resolution depends on the jet energy and is comprised in the range 0.5-1.5 mm. A somewhat more effective selection method is based on the *transverse or 3D impact parameter* which does not depend at first order on the momentum of the decaying τ (Ref. 6).

***Invariant mass:*** The τ-jet mass ($M_\tau$) is reconstructed from the momentum of the tracks in the signal cone and the energy of the cluster in the calorimeters within a certain cone $\Delta R_{jet}$ around the calorimeter jet axis. It is important to avoid double counting of particles by rejecting the calorimeter clusters which are matched to a given track. A possible un-matching condition could be that the cluster, taken for the mass calculation, must be separated from the track impact point on the calorimeter surface by a given distance $\Delta R_{tracks}$. CMS optimized cuts are $\Delta R_{jet}<0.4$ and $\Delta R_{track}>0.08$. The effectiveness of the method can be seen in Figure 5 where the invariant mass distribution for simulated signal events (τ-jets) and background QCD jet is shown for two ranges of jet transverse energy. Both experiments have developed a tag based on the invariant mass distribution; Atlas has developed a method for mass reconstruction in the context of an *energy flow* algorithm (see Ref. 5).

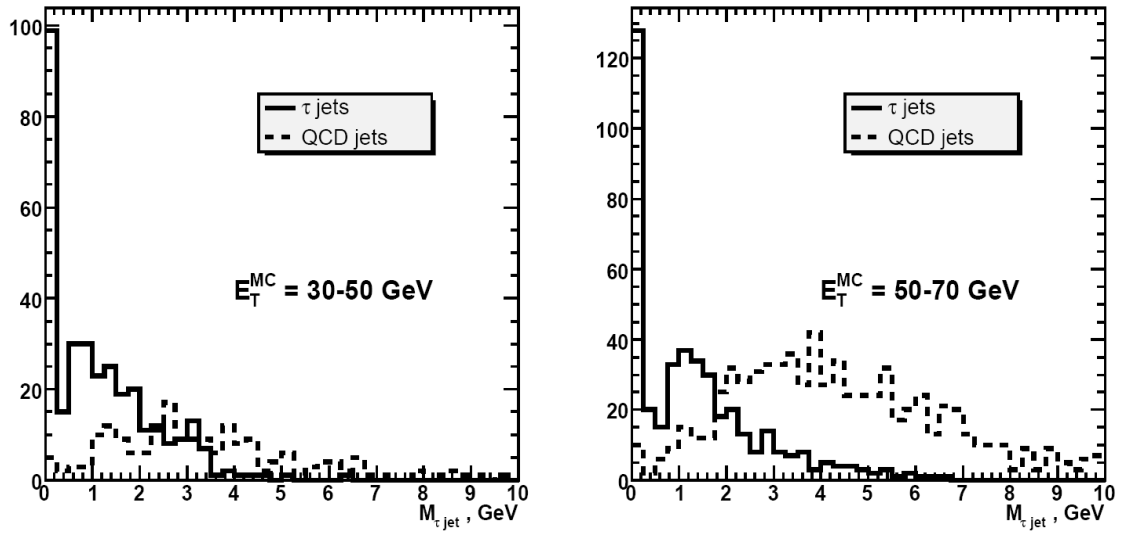

**Figure 5** CMS: Distribution of $M\tau$ for $\tau$-jets (solid line) and QCD jets (dashed line) in the interval of the true MonteCarlo transverse energy of the jet ($E_T^{MC}$) of 30-50 GeV (left plot) and 50-70 GeV (right plot)

## ATLAS SPECIFIC SELECTION

Atlas has developed two independent algorithms for $\tau$-jet selection. TauRec (Ref. 4) and Tau1P3P (Ref. 5): the former is a general purpose algorithm based on calorimeters + inner detectors information, the latter is intended for study of low mass Higgs decays.

*tauRec:* this algorithm uses calorimeter clusters as $\tau$-jet candidates. These are provided by a sliding window cluster algorithm which runs on "CaloTowers" which are the sum of all calorimeter layers summed up on a grid of $\Delta\eta \times \Delta\Phi = 0.1 \times 2\pi/64$. The $\tau$-jet are pre-selected by looking at the following quantities: isolation (cut on EM-radius and on $\Delta E_T^{12}$), number of associated charged tracks with $P_T>2$GeV and with a distance $\Delta R<0.3$ from the baricenter of the cluster, charge $\pm1$, 2D signed impact parameter significance, strip width in the so called $\eta$-strip layer of the electromagnetic calorimeter. After the pre-selection a likelihood function is built with all the previous variables including the ratio $E_T/P_T$ of the leading track. A rejection against QCD jets of $10^3$-$10^4$ is obtained (depending on the jet energy) with an efficiency of ~40% for $\tau$-jets.

*Tau1P3P:* the Tau1P3P algorithm is specialized for low mass Higgs ($m_H \sim 120$ GeV), with visible energy from hadronic $\tau$ decays the range 20-50 GeV. One or three charged tracks are required plus associated energy deposit in the calorimeters (from $\pi^{\pm}$) and additional neutral electromagnetic energy from the accompanying $\pi^0$s. The search for calorimetric energy is seeded by the direction of the leading track. The main steps of the Tau1P3P algorithm are: look for a "good" hadronic leading track ($P_T > 9$ GeV), zero or two nearby tracks with $P_T > 2$ GeV, $\Delta R$(track-direction, jet-direction) $< 0.2$, isolation in a ring $0.2<\Delta R<0.4$ around the $\tau$-jet direction. The calorimetric clusters are classified in *neutral electromagnetic*, *charged electromagnetic*, *charged hadronic* and *others* type with a simplified energy-flow method. Several additional discriminant variables are calculated by making use of the tracks and of the clusters belonging to the $\tau$-jet: invariant mass, ratio of the *charged* energy deposited in the hadronic calorimeter and the transverse track momentum, sum of transverse momenta, and some other variable. After optimizing the cuts a selection efficiency ~18.9% for $\tau$-jets coming from the decay of a Higgs boson with m=120GeV/c$^2$ is found. The jets coming from the background (QCD events) are selected with an efficiency of ~0.3% (1 prong jets) and ~1% (3 prongs jets). Better results have been obtained with a multivariate analysis (Ref. 5), which produces a higher background rejection (the QCD background selection efficiency is ~65% lower compared to the efficiency found with the "cut-based" analysis, while the signal efficiency is almost unchanged).

## CMS SPECIFIC SELECTION

CMS selection is based on the calorimetric and tracker isolation described previously. Referring to Figure 4 the optimization of the working point of the tracker isolation algorithm is done by making a scan on the value of the isolation cone ($R_i$), with the value of signal cone ($R_s$) and $R_m$ kept fixed. It is possible to reach good values of background rejection ($\varepsilon$(QCD jets) $\sim$ 4-6%) with an efficiency for $\tau$-jets of $\sim$70%. The actual signal efficiency depends on the particular physical process considered.

A number of other selection methods (impact parameter, flight path, mass reconstruction) which can be applied after the isolation criterions have been studied by the CMS collaboration. Most of these additional cuts have been already described previously in this paper (see Ref. 6 for more details). Depending on the specific channel studied, the application of these additional cuts can greatly improve the overall signal/background ratio.

## HIGH LEVEL TRIGGER

The trigger of the LHC experiments is composed by a first level trigger (LVL1) which is entirely implemented on custom hardware and performs a rapid decision based on calorimeters and muon chambers information. Subsequently a much more accurate selection is performed by software algorithms which run on large filter farms. Atlas has an intermediate level of trigger (LVL2) which is applied to the region of interest (ROI) pointed by the first level trigger, followed by the High Level Trigger (HLT) selection. CMS instead implements a one-step HLT selection just after the LVL1 trigger.

Given the expected QCD background cross-sections, the goal of the HLT for $\tau$-jets events is to reduce the rate of the QCD events of a factor $\sim 10^{-3}$ in order to select a final rate of O(10 Hz) events containing one (or two) $\tau$-jets (Ref. 1). Atlas and CMS HLT selection is generally based on algorithms very similar (or identical) to those applied for the off-line selection.

Atlas current implementation is based on LVL1 $\tau$-jet trigger which selects isolated jets and define the ROI. The LVL2 uses the information of the calorimeters and of the tracker to refine the selection with more quantities (cluster EM radius, width of energy deposition, isolation fraction, number of associated tracks). The Atlas HLT, not yet fully finalized, will be based on the tauRec selection.

CMS trigger (Ref. 6) is based on a LVL1 which selects isolated jets. The HLT must reduce the rate from O(kHz) to O(Hz), and is based on the calorimetric isolation (previously described) + tracker isolation. Given the speed of the inner pixel detector, a dedicated HLT trigger based on pixels has been also developed.

## CONCLUSIONS

The identification of the hadronic decays of tau leptons will be crucial for many analyses at the forthcoming LHC general purpose experiments. Both Atlas and CMS have developed $\tau$ tagging methods which are well advanced and have been tested with detailed Monte Carlo simulations. Given the peculiarities of $\tau$-jets a trigger chain starting from LVL1 up to the High Level Trigger is possible.

## ACKNOWLEDGMENTS

This report is based on the work of many people of the two collaborations Atlas and CMS. I would like to thank all of them for the beautiful and exciting results they have been able to produce

## REFERENCES

1. CMS Physics TDR, Volume 1, CERN-LHCC-2006-001 and Volume 2, CERN-LHCC-2006-021
2. ATLAS collaboration – "Detector and Physics performance TDR" CERN-LHCC-1999-015
3. J.Thomas et al. "Study of heavy MSSM-Higgs bosons A/H in hadronic $\tau$-decays in ATLAS", ATL-PHYS-2003-003
4. M. Heldmann, D. Cavalli, "An improved $\tau$-identification for the ATLAS experiment", ATL-PHYS-PUB-2006-008
5. E. Richter-Was, T. Szymocha, "Hadronic $\tau$ identification with track based approach", ATL-PHYS-PUB-2005-005
6. G. Bagliesi et al, "Tau jet reconstruction and tagging at High Level Trigger and off-line", CMS-NOTE 2006-028